\documentclass[prl,twocolumn,showpacs,superscriptaddress,amsfonts,amsmath,floatfix]{revtex4-1}

\usepackage{graphicx}
\usepackage{bbold}
\usepackage{color}

\begin{document}


\title{The Harper-Hofstadter Hamiltonian and conical diffraction 
in photonic lattices with grating assisted tunneling}

\author{Tena Dub\v{c}ek}
\affiliation{Department of Physics, University of Zagreb, Bijeni\v{c}ka c. 32, 10000 Zagreb, Croatia}
\author{Karlo Lelas}
\affiliation{Faculty of Textile Technology, University of Zagreb, 
Prilaz baruna Filipovi\'{c}a 28a, 10000 Zagreb, Croatia}
\author{Dario Juki\'{c}}
\affiliation{Department of Physics, University of Zagreb, Bijeni\v{c}ka c. 32, 10000 Zagreb, Croatia}
\affiliation{Max Planck Institute for the Physics of Complex Systems, N\"{o}thnitzer Str.
38, 01187 Dresden, Germany}
\author{Robert Pezer}
\affiliation{Faculty of Metallurgy, University of Zagreb, Aleja narodnih heroja 3, 
HR 44103 Sisak, Croatia}
\author{Marin Solja\v{c}i\'{c}}
\affiliation{Department of Physics, Massachusetts Institute of Technology, Cambridge, Massachusetts 02139, USA}
\author{Hrvoje Buljan}
\affiliation{Department of Physics, University of Zagreb, Bijeni\v{c}ka c. 32, 10000 Zagreb, Croatia}
\email{hbuljan@phy.hr}

\date{\today}

\begin{abstract}
We introduce a grating assisted tunneling scheme for tunable synthetic 
magnetic fields in photonic lattices, which can be implemented 
at optical frequencies in optically induced one- and two-dimensional dielectric photonic lattices. 
We demonstrate a conical diffraction pattern in particular realization 
of these lattices which possess Dirac points in $k$-space, 
as a signature of the synthetic magnetic fields. 
The two-dimensional photonic lattice with grating assisted tunneling constitutes the
realization of the Harper-Hofstadter Hamiltonian. 
\end{abstract}

\maketitle

Synthetic magnetism for photons is a unique tool for the manipulation and 
control of light, and for the design of novel topological phases and 
states in photonics~\cite{Haldane2008,Raghu2008,Wang2008,Wang2009,
Rechstman2013,Hafezi2013,Hafezi2011,Umucalilar2011,Fang2012,
Rechtsman2013a,Zeuner2012,Kraus2012,Khanikaev2013,Lu2013,Lu2014}. 
Topological photonics is a rapidly growing 
field~\cite{Lu2014}, advancing in parallel to analogous efforts in 
ultracold atomic gases~\cite{Dal2011,Goldman2014}, inspired by the 
development of topological insulators in condensed matter 
physics~\cite{HasanReview2010}. 
One motivating aspect of topological photonic 
systems is the existence of unidirectional backscattering immune 
states~\cite{Haldane2008,Raghu2008,Wang2008,Wang2009}, 
which are robust to imperfections, and thus may serve as novel 
waveguides and for building integrated photonic devices.  
The first experimental observations of such edge states were 
in the microwave domain, in magneto-optical photonic 
crystals~\cite{Wang2009}, theoretically proposed in
Refs.~\cite{Haldane2008,Raghu2008,Wang2008}. 
In the optical domain, imaging of topological edge states was  
reported in Floquet topological insulators, implemented in 
modulated honeycomb photonic lattices~\cite{Rechstman2013}, and in the 
two-dimensional array of coupled optical-ring resonators~\cite{Hafezi2013}. 
The strategies for obtaining synthetic magnetic/gauge fields and 
topological phases for optical photons are closely related to the system at hand. 
In systems of coupled optical resonators, the strategy is to tune the 
phase of the tunneling between coupled cavities~\cite{Hafezi2011,Umucalilar2011,
Fang2012,Hafezi2013}; for example, 
by using link resonators of different length~\cite{Hafezi2011, Hafezi2013}
or time-modulation of the coupling~\cite{Fang2012}.
Photonic topological insulators were also proposed in superlattices of metamaterials 
with strong magneto-electric coupling~\cite{Khanikaev2013}. 
In 2D photonic lattices (waveguide arrays), pseudomagnetic fields have been 
demonstrated by inducing 'strain' in optical graphene~\cite{Rechtsman2013a}.
By modulating 1D photonic lattices along the propagation axis, 
one can choose the sign of the hopping parameter between neighboring 
sites~\cite{Zeuner2012}, whereas topological states were 
achieved using 1D photonic quasicrystals~\cite{Kraus2012}. 
However, even though photonic lattices possess a great potential for exploring 
synthetic magnetism and topological effects, a viable scheme for arbitrary 
design of the phases of the complex tunneling matrix elements still needs to be developed. 
Here we introduce one such scheme termed grating assisted tunneling, 
and propose its implementation in optically induced photonic 
lattices~\cite{Nicos2002,Fleischer2003a,Fleischer2003,Neshev2003,Fleischer2005}. 
We demonstrate the conical diffraction pattern \cite{Moti2007,Berry2004} in particular realizations 
of one- (1D) and two-dimensional (2D) square photonic lattices with 
grating assisted tunneling. The latter constitutes demonstration of the 
the ('single particle') Harper-Hofstadter Hamiltonian (HHH)
\cite{Harper1955,Hofstadter1976} in these systems.

The proposed method is inspired by the so called laser assisted tunneling 
scheme, which was implemented in optical lattices with ultracold 
atoms~\cite{Aidelsburger2011,Miyake2013,Aidelsburger2013}. The scheme 
is based on theoretical proposals in Refs.~\cite{Jaksch2003,Kolovsky2011}, subsequently developed to 
experimentally realize the HHH~\cite{Miyake2013,Aidelsburger2013}, and 
staggered magnetic fields in optical superlattices~\cite{Aidelsburger2011}. 
Many of the results obtained with ultracold atoms are viable in photonic lattices, 
and perhaps even more of them are feasible, because heating by spontaneous emission 
that is present in ultracold atomic systems, is absent in photonic systems.

We consider the paraxial propagation of light in a photonic 
lattice defined by the index of refraction $n = n_0 + \delta n(x,y,z)$
($\delta n\ll n_0$),
where $n_0$ is the constant background index of refraction, and $\delta n(x,y,z)$ 
describes small spatial variations, which are slow along the propagation $z$-axis. 
The slowly varying amplitude of the electric field $\psi(x,y,z)$ follows the 
(continuous) Schr\"{o}dinger equation:
\begin{equation}
i\frac{\partial \psi}{\partial z}=
-\frac{1}{2k}\nabla^2 \psi-\frac{k \delta n }{n_0}\psi;
\label{sys}
\end{equation}
here, $\nabla^2=\partial^2/\partial x^2+\partial^2/\partial y^2$, and 
$k=2\pi n_0/\lambda$, where $\lambda$ is the wavelength in vacuum.
From now on we will refer to $\delta n(x,y,z)$ as the potential or index of refraction.
In the simulations we use $n_0=2.3$, corresponding to 
the systems that were used to implement the optical induction 
technique~\cite{Fleischer2003a,Fleischer2003}, and $\lambda=500$~nm.

\begin{figure*}
\centerline{
\mbox{\includegraphics[width=0.85\textwidth]{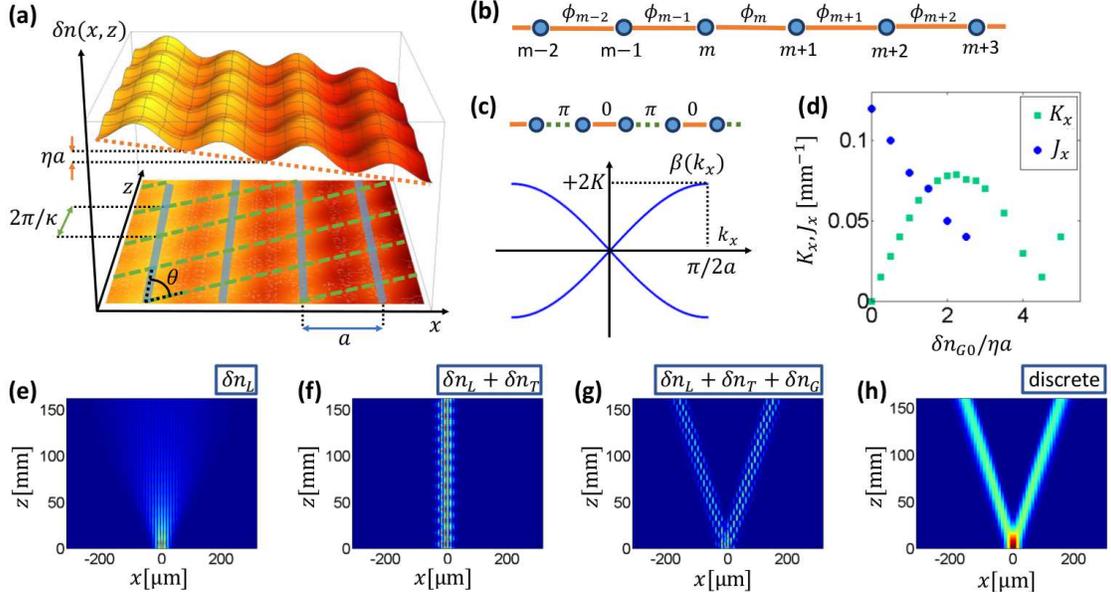}}
}
\caption{(color online) 
Illustration of the grating assisted tunneling scheme.
(a) Sketch of the spatial dependence of the index of refraction
$\delta n_{L}(x,z)=\delta n_{L}+\delta n_{T}+\delta n_{G}$, which 
creates a synthetic magnetic field in a photonic lattice. 
A 1D photonic lattice [$\delta n_{L}=\delta n_{L0}\cos^2(\pi x/a)$] 
is superimposed with a linear gradient index in the $x$ direction 
($\delta n_{T}=-\eta x$, orange dotted line), and an additional 
small grating potential [$\delta n_{G}=\delta n_{G0} \cos^2((q_x x-\kappa z) /2)$], 
at a small angle ($\theta$ is on the order of $1^{\circ}$) with respect to the $z$ axis.
In the $xz$ plane we plot the projection of $\delta n_{L}$ (blue bold lines), 
and the grating (green dashed lines). 
(b) A discrete lattice with complex tunneling matrix elements between 
sites, $K_x e^{i\phi_m}$, and spatially dependent phases $\phi_m$, 
can model the system in (a). 
(c) A particular choice of $\theta$ (see text), yields a lattice with 
$\phi_m=\pi m$, and a dispersion with a 1D Dirac cone at $k_x=0$.
(d) The amplitude of the tunneling $K_x$ as a function 
of the strength of the grating $\delta n_{G0}$ (green squares). 
The tunneling amplitude $J_x$ versus $\delta n_{G0}$ for a system 
without the tilt. 
(e,f,g) Numerical simulation of the evolution of a wavepacket that initially
excites modes close to $k_x=0$ in the continuous 1D photonic lattice. 
(e) Diffraction in a periodic photonic lattice $\delta n_{L}$. 
(f) Propagation in the tilted system, $\delta n_{L}+\delta n_{T}$, 
shows that the tunneling (diffraction) is suppressed. 
(g) The tunneling is restored by an additional grating potential with $q_x=\pi/a$. 
(h) Propagation of the wavepacket in the 
corresponding discrete model illustrated in (c).
}
\label{1Dphase}
\end{figure*}

For clarity, let us first present the grating assisted tunneling method in a 
1D photonic lattice.
If the potential is a periodic lattice, 
$\delta n_{L}(x)=\delta n_{L}(x+a)$, where $a$ is the lattice constant, 
and if the lattice is sufficiently 'deep', the propagation of light can be approximated 
by using a discrete Schr\"{o}dinger equation~\cite{Fleischer2005,Demetri2003}:
\begin{equation}
i\frac{d \psi_{m}}{d z} = 
- (J_x\psi_{m-1}+J_x\psi_{m+1}),
\label{model}
\end{equation}
where $J_{x}$ quantifies the tunneling between adjacent waveguides and 
$\psi_{m}(z)$ is the amplitude at the $m$th lattice site, i.e., waveguide. 
To create a synthetic magnetic field, we adopt the strategy to tune the 
phase of the tunneling between lattice sites. For this 1D lattice, 
we seek a scheme which effectively renormalizes $J_{x}$ to get $K_x \exp(i\phi_m)$, where 
$\phi_m$ denotes the phase for tunneling from site $m$ to site $m+1$. 
The scheme is illustrated in Fig. \ref{1Dphase}.
In Fig. \ref{1Dphase}(a) we show the potential $\delta n(x,z)$, 
which can be modeled as a discrete lattice with complex tunneling 
parameters $K_x \exp(i\phi_m)$ shown in Fig.~\ref{1Dphase}(b).

Let us gradually explain the idea behind the variation of the index of refraction as in 
Fig.~\ref{1Dphase}(a). 
In Fig.~\ref{1Dphase}(e), we show the propagation of intensity in a continuous 
1D model in the lattice potential $\delta n_{L}(x)=\delta n_{L0}\cos^2(\pi x/a)$, 
with $\psi(x,0)=\sqrt{I} e^{-x^2/(3 a)^2}$; $\delta n_{L0}=4\times 10^{-4}$, $a=10$~$\mu$m. 
We see the usual diffraction pattern for a spatially broad excitation 
covering several lattice sites~\cite{Fleischer2005}. 
Next, suppose that we introduce a linear gradient of index of 
refraction along the $x$ direction $\delta n_{T}(x)=-\eta x$ in addition 
to the lattice potential, such that $\delta n=\delta n_{L}(x)+\delta n_{T}(x)$. 
For a sufficiently large tilt, the tunneling is {\em suppressed}. 
This can be seen from Fig. \ref{1Dphase}(f) which shows the 
propagation of intensity in a tilted potential with $\eta=0.1\delta n_{L0}/a$; 
the tilt should be smaller than the gap between the first two bands. 
Finally, let us introduce an additional small grating potential at a small angle $\theta$ 
with respect to the $z$ axis, $\delta n_{G}(x,z)=\delta n_{G0} \cos^2((q_x x-\kappa z) /2)$, 
such that $\delta n (x,z)=\delta n_{L}(x)+\delta n_{T}(x)+\delta n_{G}(x,z)$.
This total potential $\delta n (x,z)$ is illustrated in Fig. \ref{1Dphase}(a). 
The 'frequency' $\kappa$ is determined by the angle $\theta$ of the grating with 
respect to the $z$-axis, which is chosen such that $\kappa = \eta a k/n_0$ and
the grating forms a $z$-dependent perturbation 
resonant with the index offset between neighboring lattice sites $\eta a$
(Fig. \ref{1Dphase}(a)). 
The grating {\em restores} the tunneling along the $x$-axis, hence the 
term {\em grating assisted tunneling}. 
Restored tunneling is seen in Fig. \ref{1Dphase}(g) which shows 
diffraction for identical initial conditions as in Figs. \ref{1Dphase}(e) and (f); 
the grating parameters are $q_x=\pi/a$ and $\delta n_{G0}=0.1\delta n_{L0}$.

However, the diffraction pattern is drastically changed. In order to interpret it, 
we point out that, for resonant tunneling where $\kappa = \eta a k/n_0$ 
and a sufficiently large tilt ($J\ll\eta$), $z$-averaging over the rapidly 
oscillating terms shows that the system can be modeled by 
an effective discrete Schr\"{o}dinger equation (e.g., see~\cite{Miyake2013} 
for ultracold atoms):
\begin{equation}
i\frac{d \psi_{m}}{d z}=
- (K_x e^{i\phi_{m-1}} \psi_{m-1} + K_x e^{-i\phi_{m}}\psi_{m+1} ),
\label{1DHH}
\end{equation}
where $\phi_{m}={\bf q}\cdot {\bf R}_{m}=q_x m a$.
In Fig. \ref{1Dphase}(g) we used $q_x=\pi/a$, i.e., $\phi_{m}=m\pi$. 
Such a discrete lattice is illustrated in Fig. \ref{1Dphase}(c); 
its dispersion having a 1D Dirac cone at $k_x=0$. 
For a wavepacket that initially excites modes close 
to $k_x=0$, the diffraction in the discrete model (\ref{1DHH}) yields the so-called 
(1D) conical diffraction pattern \cite{Zeuner2012,Moti2007}, as illustrated in Fig.~\ref{1Dphase}(h). 
The initial conditions for propagation in the discrete model 
corresponds to the initial conditions in the continuous system, 
$\psi_m(0)=\sqrt{I} e^{-(m/3)^2}$, and $K_x=0.053$~mm$^{-1}$. 
Thus, we interpret the diffraction pattern in Fig. \ref{1Dphase}(g) as 
1D conical diffraction, a signature of the discrete model depicted 
in Fig. \ref{1Dphase}(c). 
A comparison of the discrete [Fig.~\ref{1Dphase}(h)] and the realistic 
continuous model [Fig. \ref{1Dphase}(g)], clearly shows that we can use 
grating assisted tunneling to tune the phases of the tunneling parameters
in the discrete Schr\"{o}dinger equation, 
thereby realizing synthetic magnetic fields.

Before proceeding to 2D systems, we discuss the amplitude 
of the tunneling matrix elements $K_x$  as a function
of the strength of the grating $\delta n_{G0}$. 
Figure \ref{1Dphase}(d) shows $K_x$ (green squares) and $J_x$ (blue circles) 
versus $\delta n_{G0}$, where $J_x$ corresponds to the potential
which includes the lattice and the grating, but no tilt.
The amplitudes $J_x$ and $K_x$ are obtained 
by comparing the diffraction pattern of the discrete with the
continuous model, and adjusting $J_x$ and $K_x$ until the two patterns coincide, 
as in Figs.~\ref{1Dphase}(g) and (h). Our results in 
Fig.~\ref{1Dphase}(d) are in agreement with those in ultracold atoms 
[e.g., see Fig. 3(a) in Ref.~\cite{Miyake2013}].

The extension of the scheme to 2D lattices is straightforward. 
We consider a square lattice, $\delta n_{L}=\delta n_{L0}(\cos^2(\pi x/a)+\cos^2(\pi y/a))$, 
the tilt in the $x$ direction, $\delta n_{T}(x)=-\eta x$, 
and the grating which has the form 
$\delta n_{G}(x,y,z)=\delta n_{G0} \cos^2((q_x x + q_y y - \kappa z)/2)$.
Propagation of light in the total potential $\delta n(x,y,z)=
\delta n_{L}(x,y)+\delta n_{T}(x)+\delta n_{G}(x,y,z)$
can be modeled by the discrete Schr\"{o}dinger equation
(the derivation is equivalent to that in Ref.~\cite{Miyake2013} for ultracold atoms):
\begin{eqnarray}
i\frac{d \psi_{m,n}}{d z}
& = &
- (K_x e^{i\phi_{m-1,n}} \psi_{m-1,n} + K_x e^{-i\phi_{m,n}}\psi_{m+1,n}
\nonumber \\
& + & J_y \psi_{m,n-1} + J_y \psi_{m,n+1}),
\label{HH}
\end{eqnarray}
where $\phi_{m,n}={\bf q}\cdot {\bf R}_{m,n}=q_x m a+q_y n a$. 
Note that the tunneling along $y$ does not yield a phase because 
there is no tilt in the $y$ direction; the tunneling amplitude along $y$ 
depends on the depth of the grating, as illustrated in Fig. \ref{1Dphase}(d) 
with blue circles. 

\begin{figure}
\centerline{
\mbox{\includegraphics[width=0.50\textwidth]{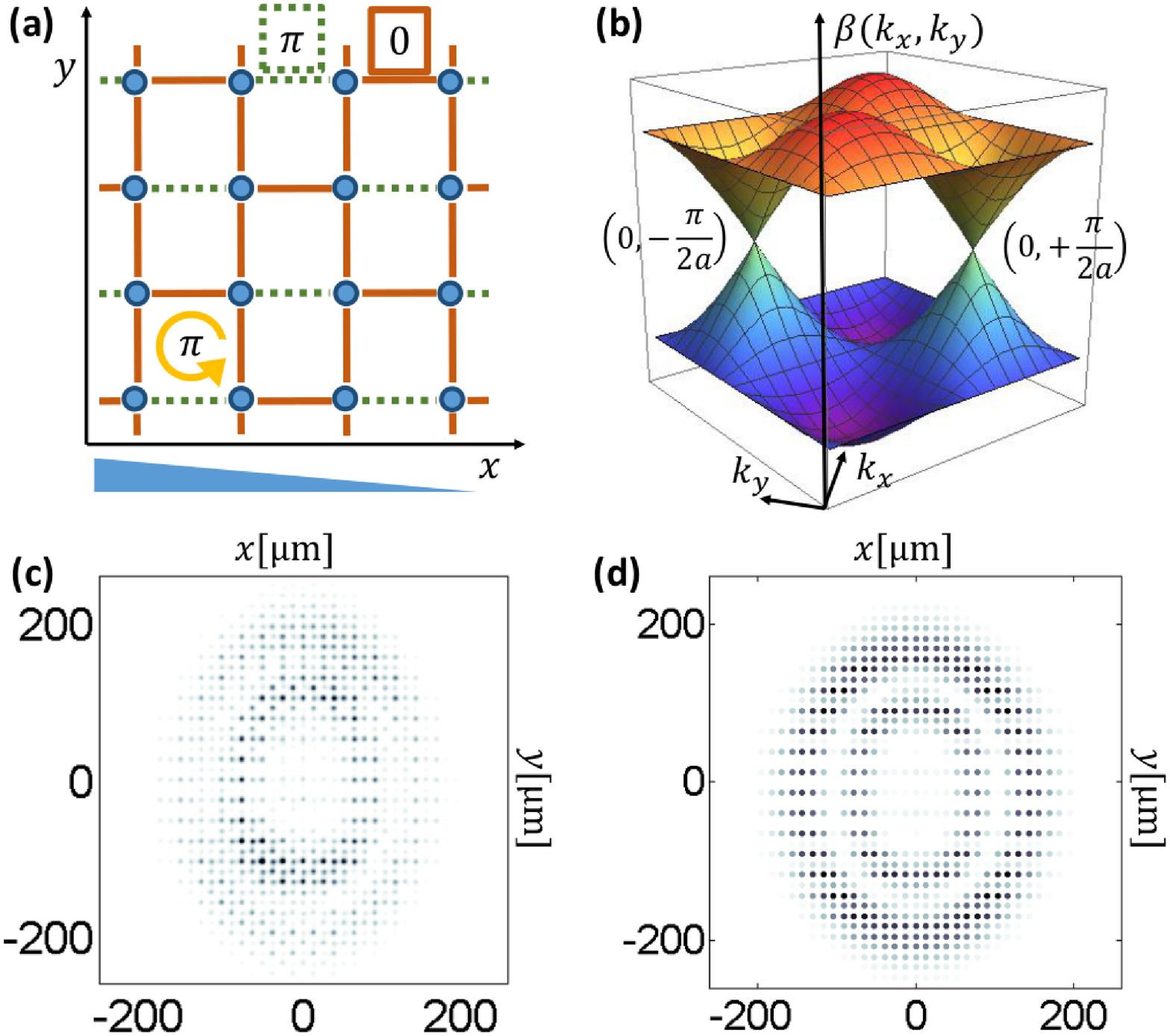}}
}
\caption{(color online) 
Grating assisted tunneling and conical diffraction in a square photonic lattice.
(a) Sketch of the 2D lattice with grating assisted tunneling along the $x$ direction.
The resulting nontrivial hopping phases $\pi$ and $0$ are denoted with dashed and 
solid lines, respectively. A wavepacket that makes one loop around 
the plaquette accumulates the phase $\pi$. 
(b) The lattice possesses two 2D Dirac cones at $(k_x,k_y)=(0,\pm\pi/2a)$ in the 
dispersion $\beta(k_x,k_y)$, where $\beta$ is the propagation constant.
(c) Intensity of a beam, which initially excites modes in the vicinity of the 
Dirac points, after propagation for $z=162$~mm. The intensity has 
two concentric rings corresponding to the conical diffraction pattern. 
(d) Simulation of the diffraction pattern in the discrete model
corresponding to the lattice in (a), which also exhibits conical diffraction
(see text for details).
}
\label{2Dphase}
\end{figure}

In order to demonstrate that the propagation of light in the 
continuous 2D potential $\delta n(x,y,z)$ is indeed equivalent to the 
dynamics of Eq. (\ref{HH}), we compare propagation 
in the discrete model (\ref{HH}), with that of the continuous equation (\ref{sys}). 
The lattice parameters are $\delta n_{L0}=4\times 10^{-4}$, $a=13$~$\mu$m; 
the tilt is given by $\eta=0.1\delta n_{L0}/a$; 
the grating is defined by $\delta n_{G0}=0.17\delta n_{L0}$ and 
$q_x=-q_y=\pi/a$, which yields $\phi_{m,n}=(m-n)\pi$, and $\kappa$ is chosen to 
yield resonant tunneling. 
The discrete lattice which corresponds to this choice of phases is 
illustrated in Figure \ref{2Dphase}(a). It has two bands,
$\beta=\pm 2 \sqrt{K_x^2 \sin^2(k_xa) + J_y^2 \cos^2(k_ya)}$ ($\beta$ is the propagation constant),
touching at two 2D Dirac points at $(k_x,k_y)=(0,\pm \pi/2a)$ in the Brillouin zone~\cite{MiyakePhD2013},
as depicted in Fig.~\ref{2Dphase}(b). 
Suppose that the incoming beam at $z=0$,  
excites the modes which are in the vicinity of these two Dirac points. 
Around these points, for a given ${\hat{\bf k}}={\bf k}/k$, the group velocity, 
$\nabla_{\bf k} \beta (k_x,k_y)$, is constant. 
Thus, the beam will undergo conical 
diffraction, which has been thoroughly addressed with Dirac points in 
honeycomb optical lattices~\cite{Moti2007}. To demonstrate this effect, we
consider the propagation of a beam with the initial profile given by 
$\psi(x,y,0)=\sqrt{I} \cos(y\pi/2a) \exp(-(x/3a)^2-(y/3a)^2)$ in the 2D potential $\delta n(x,y,z)$
(the cosine term ensures that we excite modes close to the two Dirac points). 
In Fig.~\ref{2Dphase}(c) we show this beam after propagation for $z=162$~mm. 
The two concentric rings are a clear evidence of conical diffraction~\cite{Moti2007,Berry2004}. 
We also compare this with the propagation in the discrete model (\ref{HH}), 
with tunneling phases plotted in Fig.~\ref{2Dphase}(a); $K_x=0.11$~mm$^{-1}$, $J_y=0.14$~mm$^{-1}$. 
The results of the propagation in the discrete model are shown in 
Fig.~\ref{2Dphase}(d). It is evident that for our choice of 
the tilt and the grating, we have effectively realized the lattice 
plotted in~\ref{2Dphase}(a). This is in fact a realization of the HHH for 
$\alpha=1/2$, where $\alpha$ is the flux per plaquette  
in units of the flux quantum~\cite{Miyake2013,Aidelsburger2013}.

For the experimental implementation of the scheme, we propose the so-called 
optical induction technique in photosensitive materials, 
which can be implemented in 
photorefractives~\cite{Nicos2002,Fleischer2003a,Fleischer2003,Neshev2003,Fleischer2005}. 
In these systems, both the lattice $\delta n_{L}(x,y)$ and the grating potential 
$\delta n_{G}(x,y,z)$ can be obtained in a straightforward fashion by using 
interference of plane waves in the medium~\cite{Fleischer2005}.
The lattice constant $a$, the grating parameter ${\bf q}$, 
and hence the hopping phase $\phi_{m,n}={\bf q}\cdot {\bf R}_{m,n}$, are tunable 
by changing the angle between the interfering beams. 
This could enable manipulation of the phases of the tunneling matrix 
elements in real time~\cite{Fleischer2003a}.

The most challenging part of the implementation appears to be 
creation of the linear tilt potential. To observe the 
conical diffraction effects, one needs a linear gradient over the
$\sim 20$ lattice sites, which implies a total index tilt 
$\Delta n = 20 \eta a \sim 20 \times 0.1 \delta n_{L0}\sim 8 \times 10^{-4}$. 
In principle, the linear tilt could be achieved by using a spatial light modulator. 
Another possibility which could create a linear tilt is to use crystals 
with linear dependence of the index of refraction on temperature, 
and a temperature gradient across the crystal.

However, it should be emphasized that if one uses a superlattice for 
$\delta n_L$, rather than the linear tilt potential, as in Ref.~\cite{Aidelsburger2011} 
for ultracold atomic gases, the grating assisted tunneling scheme 
proposed here would straightforwardly yield {\em staggered} synthetic magnetic 
fields for photons. The achievement of such staggered fields is 
straightforward with optically induced lattices proposed here.

Before closing, we emphasize that the spatial phase between the periodic 
lattice potential $\delta n_L$ and the grating $\delta n_G$ is a relevant parameter. 
This point has recently been examined in detail in the context of ultracold atoms~\cite{Tarallo2012}. 
To translate it to our system, suppose that in our 1D system, we shift the grating along the $x$-direction,
such that $\delta n_{G}=\delta n_{G0} \cos^2((q_x x-\kappa z+\xi) /2)$.
This simply shifts the phases in the discrete lattice [shown in Fig.~\ref{1Dphase}(c)] such that 
$\phi_m$ is replaced by $\phi_m+\xi$, which moves the 1D Dirac point in $k_x$-space 
by $\Delta k_x=\xi/a$. We have numerically verified that this indeed happens.

In conclusion, we have introduced a grating assisted tunneling scheme 
for tunable synthetic magnetic fields in photonic lattices, and proposed its 
implementation at optical frequencies in optically induced one- and two-dimensional 
dielectric photonic lattices. 
As a signature of the synthetic magnetic fields, we have 
demonstrated the conical diffraction pattern in particular realization 
of these lattices [shown in Fig.~\ref{1Dphase}(c) and Fig.~\ref{2Dphase}(a)]. 
The two-dimensional photonic lattice with grating assisted tunneling constitutes 
the realization of the HHH. 
The scheme is well suited for realization of staggered synthetic magnetic fields
in photorefracives. 
We envision that this proposal will open the way to other studies 
of light propagation in photonic lattices with complex tunneling matrix elements,
including nonlinear propagation, solitons and instabilities in synthetic magnetic fields,
and creation of synthetic dimensions in photonic lattices~\cite{Jukic2013} by employing synthetic fields.

This work was supported by the Unity through Knowledge Fund (UKF Grant No. 5/13).
Work by MS (analysis and proof-reading of the manuscript) was supported
as a part of S3TEC, and EFRC, funded by U.S. DOE,
under Award Number DE-SC0001299 / DE-FG02-09ER46577.
We are grateful to Colin J. Kennedy, Wolfgang Ketterle, and Ling Lu for useful 
conversations. 

{\em Note added.} A few days before this paper was submitted to arXiv, a paper by a 
different group~\cite{Mukherjee2015} appeared on the arXiv. 
It is also inspired by laser assisted tunneling in cold gases, and experimentally 
demonstrated modulation-assisted tunneling in laser-fabricated photonic Wannier-Stark ladders.


\end{document}